\documentclass[prd,aps,twocolumn,nofootinbib]{revtex4}
\usepackage[dvipdfm]{graphicx}
\usepackage{amssymb}
\usepackage{amsmath,color}
\usepackage{epsfig}
\newcommand{\ba}{\begin{eqnarray}}
\newcommand{\ea}{\end{eqnarray}}
\newcommand{\be}{\begin{equation}}
\newcommand{\ee}{\end{equation}}
\newcommand{\al}{\alpha}
\newcommand{\bt}{\beta}
\newcommand{\ga}{\gamma}
\newcommand{\ta}{\theta}

\newcommand{\da}{\delta}
\newcommand{\la}{\lambda}
\newcommand{\ka}{\kappa}

\newcommand{\en}{\epsilon}

\newcommand{\Ga}{\Gamma}

\newcommand{\Da}{\Delta}
\newcommand{\Oa}{\Omega}
\newcommand{\La}{\Lambda}

\newcommand{\cP}{{\cal P}}

\newcommand{\cO}{{\cal O}}

\newcommand{\cN}{{\cal N}}

\newcommand{\p}{\partial}

\newcommand{\ra}{\rightarrow}
\newcommand{\Ra}{\Rightarrow}

\newcommand{\LF}{\left(}
\newcommand{\RF}{\right)}
\newcommand{\LT}{\left[}
\newcommand{\RT}{\right]}

\newcommand{\ke}{k'}

\newcommand{\2}{\frac{1}{2}}

\newcommand{\4}{\frac{1}{4}}

\newcommand{\mx}{\mbox}
\newcommand{\mt}{\mathtt}

\newcommand{\for}{\mx{ for }}

\newcommand{\ie}{{\it i.e.\ }}

\begin{document}

\title{Wiggles in the cosmic microwave background radiation: echoes from non-singular cyclic-inflation}
\author{Tirthabir Biswas$^{a,b}$}
\author{Anupam Mazumdar$^b$}
\author{Arman Shafieloo$^c$}
\affiliation{
{\it $^a$Department of Physics,
St. Cloud State University, St. Cloud, MN 56301}\\
{\it $^b$Department of Physics,
Loyola University, New Orleans, LA 70118}\\
{\it $^c$ Physics Department, Lancaster University, Lancaster, LA1 4YB, UK}\\
{\it $^d$ Physics Department, University of Oxford, Oxford, OX1 3NP, UK}
}
\date{\today}
\begin{abstract}
In this paper we consider a unique model of inflation where the universe undergoes rapid asymmetric  oscillations, each cycle
lasting $\sim 10^{6}$ Planck times. Over many-many cycles the space-time expands to mimic the  standard inflationary scenario. Moreover,  these rapid oscillations  leave a distinctive periodic signature in $\ln k$ in the primordial power spectrum, where $k$ denotes the comoving scale. Although the cyclic-inflation model contains additional parameters  as compared to the standard power-law  spectrum,  the improvement to the fit  of the 7-year WMAP data is significant.
\end{abstract}

\maketitle
\section{Introduction}
Primordial inflation has been very successful in explaining the perturbations in the cosmic microwave background (CMB) radiation and the large scale structures of the universe~\cite{wmap7}, for a recent review see~\cite{RM}. However inflation
has some outstanding problems, in particular it doesn't encode a non-singular geodesically complete evolution~\cite{Linde1}.
We will present a unique singularity free geodesically complete realization of inflation in the context of cyclic cosmologies.

In cyclic cosmological models, rather than having  a singular {\it beginning of time}, our universe can be made {\it eternal} in both past and future~\cite{tolman,narlikar,Starobinsky,ekcyclic,barrow,phantom,peter,emergent,BM,saridakis}. However, in most cyclic scenarios the effort has been to produce primodial fluctuations within a single long cycle~\footnote{Recently, there have been attempts to calculate how perturbations can evolve through various cycles~\cite{robert}.}. Although there have been progress~\cite{justin,BMS,matter}, this has  proved  rather challenging in comparison to the success which inflation enjoys in explaining the observed near scale invariant perturbations in the CMB.

In this paper we provide a simple alternative to  the standard inflation and cyclic universe scenarios in the form of a cyclic-inflation model which tries to incorporate the successes of both; our model  includes a non-singular cyclic phase of evolution where in every cycle the universe contracts a little less than it expands leading to an overall  growth. In fact, over many-many cycles the space-time resembles that of inflation. Thus, in close analogy with inflation, we can explain how  the seed perturbations generated at much higher energy densities can be stretched to the observable scales,  and
why the spectrum is nearly scale free. Additionally, the model leaves distinct signatures of the rapid oscillations the universe undergoes by modifying the power spectrum with periodic signatures. Last but not the least, it turns out that the cyclic-inflationary phase requires a negative potential energy, but the universe can gracefully exit to a positive potential region marking the end of the inflationary phase and the onset of standard radiation dominated era. Thus our model may provide a way of reaching a positive energy vacuum from a plethora of negative energy vacua in the string landscape~\cite{landscape}. Finally, Tolman's entropy problem (which is equivalent to the problem of geodesic incompleteness in our model) can be naturally addressed by including a pre-inflationary emergent phase where the scale-factor starts to oscillate periodically as we approach past infinity, the size of the universe never becomes vanishingly small~\cite{emergent}.

Let us consider a simple  cyclic inflation model, where the universe is mostly dominated by radiation, and the cycles are (approximately) periodic in energy densities. This follows if firstly, we assume  that  quantum gravitational effects trigger a bounce whenever some critical Planckian energy density is reached. Secondly, we need a $-$ve cosmological constant (CC), $-\La$, which ensures that  the universe turns around and starts to re-collapse once the radiation energy density dilutes and becomes equal to $\La$. Thus contrary to common expectations, in the presence of matter the universe does not get stuck in an AdS vacua~\cite{linde,BM}, but rather starts to cycle. These cycles are typically short, the time period, $\tau=\al{M_p/ \sqrt{\La}}$, where $\al\sim \cO(1)$ and $M_p=2.4\times 10^{18}$~GeV~\cite{BM}. We shall show that in order to obtain the correct amplitude of CMB fluctuations we will require $\La$ to be close to the conventional string/GUT scale,  $\La^{1/4} \sim 10^{-3}M_p$, so that  $\tau\sim 10^6 M_p^{-1}$.

Now, provided there is exchange of energy between radiation and some other forms of matter, then, as a natural consequence of the second law of thermodynamics,  one expects the cycles to be asymmetric. The total entropy in the universe increases monotonically, and  by the same factor in every cycle: $S_{n+1}/S_n\equiv 1+3\ka$. Since entropy is proportional to the volume this means that if, for instance, we track the scale factor at the bounce point of consecutive  cycles then
\be
\label{kappa}
{a_{n+1}}/{a_n}= 1+\ka\for \ka\ll1
\ee
While the above scenario can be realized in many different ways, here we will consider a simple toy model with two species, massless radiation and some massive particle which interact with each other. It is clear from (\ref{kappa}) that, over many asymmetric cycles the evolution of the universe looks very similar to that of ordinary inflation with an average Hubble expansion rate $    H_{\mt{av}}={\ka/\tau}$. We will see later that  this ``cyclic inflationary'' phase can indeed address the usual cosmological puzzles such as isotropy, horizon, flatness and homogeneity.

\begin{figure}[htbp]
\begin{center}
\includegraphics[width=0.35\textwidth,angle=0,scale=0.8]{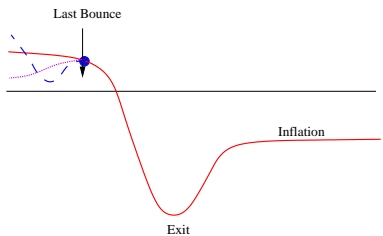}
\end{center}
\caption{Typical potentials: In the positive energy side three different possibilities are consistent with our model.
  \label{fig:potential}}
\end{figure}

What about the spectrum of the primordial fluctuations? To match the COBE normalization,  the power-spectrum associated with metric fluctuations must be given by: $\cP_{\Phi}\sim 10^{-10}$. Now in general, since matter couples very weekly to gravity, in the sub-Hubble phase (when the wavelength of a given comoving mode is smaller than the cosmological time-scale), when the metric fluctuations are generated from the matter fluctuations the amplitude is suppressed by the Planck scale. Typically we have
\be
\cP_{\Phi}\propto k^3\Phi_k^2\sim {\rho/ M_p^4}\sim 10^{-10}\,.
\label{amplitude}
\ee
Once the  wavelength becomes larger than the cosmological time scale, the metric fluctuations effectively freeze at the value of
$\rho$  when the particular mode crosses the Hubble radius.
This intuitive picture will essentially let us argue why the perturbations in our model will have  a near scale invariant spectrum with a distinctive periodic feature (in $\ln k$)  that, in fact, provides a significant improvement in the WMAP 7-yr fit, and hence may be detectable in the future experiments. The two most important parameters governing the physics are  $\La$ and $\ka$.  While the former determines the amplitude of fluctuations, the latter characterizes the wiggles on top of the near scale-invariant spectrum.

Finally, let us discuss the graceful exit problem in this inflationary scenario. If we are stuck in a $-$ve CC, then the above inflationary phase persists forever and one can never obtain an universe like ours.
Fortunately, one can exit the inflationary phase if instead of a negative cosmological constant we have a dynamical scalar field whose potential {\it interpolates} between a negative and a positive cosmological constant as depicted in Fig.~\ref{fig:potential}. Since $V(\phi)\ra -\La$ as $\phi\ra \infty$, we can realize the inflationary phase, but the scalar field keeps rolling towards smaller $\phi$ and eventually there comes a (last) cycle, in this last contraction phase the scalar field gains enough energy to zoom through the minimum and reach the $+$ve CC phase.

The paper is organized as follows: In the following section~\ref{sec:background}, we describe our toy model and the background evolution which mimics the inflationary space-time. Next, in section~\ref{sec:perturbations}, we explain why we expect to get a nearly scale-invariant spectrum with characteristic small wiggles in our model. We also try to fit our model with the WMAP data and report our findings. In~\ref{sec:conclusions}, we end with a discussion of the standard cosmological puzzles in the context of our model and an outlook towards future research directions.
\section{The Model and Background Evolution}\label{sec:background}
\subsection{The Cycles}
Let us start by looking at  a simple toy model where we have a negative cosmological constant, and the ``matter content'' of the universe  consists of a single non-relativistic species along with relativistic degrees of freedom. The scenario we will present here is very similar to what was considered in~\cite{BM}, except that we will have a  radiation dominated universe, where as in~\cite{BM} the dominant contribution to energy density always came from the non-relativistic species. The advantage of having a radiation dominated universe is that the model can then easily avoid the Black-hole over-production problem, common in cyclic models, as the Jeans length is always large and comparable to the Hubble length. Most of the analysis in the present radiation dominated scenario closely parallels what was carried out in~\cite{BM}, only some numerical factors change. Hence, we will keep this part of the discussion brief, and refer the readers to the appendix for details.

To obtain the cyclic evolotuion, we are going to make a couple of assumptions: Firstly, we will assume that during contraction if a critical Planckian energy density, $\rho_b$,  is reached, the universe bounces back nonsingularly to a phase of expansion~\footnote{Although several different bouncing models  have been considered in the literature, in most models a bounce occurs when the energy density reaches close to the Planck density, see for instance~\cite{BMS,BKM,loop,phantom}}. We will see later that the details of the mechanism of the bounce is not particularly important for our model. Secondly we are going to demand that the relativistic and the
non-relativistic species remain in thermal equilibrium above a certain critical temperature, $T_c$, but  below this temperature
the massive non-relativistic degrees of freedom fall out of equilibrium, and consequently
when they decay into radiation  thermal entropy is generated. We will see shortly that this is the crucial ingredient that makes our cycles slightly asymmetric and causes the universe to ``effectively'' inflate.

In the non-thermal phase the Hubble equation  reads:
\be
H^2={T_c^4\over 3M_p^2}\LF {\Oa_r\over a^4}+{\Oa_m\over a^3}-{\La\over T_c^4}\RF\,,
\label{hubble}
\ee
where $-\La\equiv -M_s^4$ is the negative cosmological constant, and  the $\Oa$'s are related to the energy densities via
\be
\rho_m=T_c^4{\Oa_m\over a^3}~~~\mx{ and }~~~\rho_r=T_c^4{\Oa_r\over a^4}\,.
\label{energydensities}
\ee
For definiteness, we consider a compact  universe~\footnote{For an open or a flat universe one
just has to rephrase all the arguments in terms of entropy density rather than the total entropy and volume
of the universe.} with a volume $V\equiv T_c^{-3}a^3$. Let us define the ratio of the equilibrium energy densities
of the  non relativistic (massive) and relativistic (massless)   species  at the transition point, $T=T_c$, to be given by
\be
\mu\equiv{\rho_{m,c}\over \rho_{r,c}}\,.
\ee
We will here be interested in scenarios where $\mu\ll1$, so that to a good approximation the universe is always radiation dominated.
In order to understand the dynamics, it is instructive to first look into the evolution when non-relativistic and relativistic
species are non-interacting.  Then $\Omega$'s would just be constants yielding a Friedmann-Robertson-Walker (FRW) cosmology, and we will have a cyclic universe scenario. Let us start tracking the evolution just as the universe enters the non-thermal phase at the transition temperature, $T_c$, during the expanding branch. As the universe expands further, the temperature decreases until $T\sim \La^{1\over 4}$ at which point the universe turns around due to the presence of the negative cosmological constant. As it contracts  the temperature rises. Once the temperature crosses $T_c$, the universe enters the thermal (still contracting) phase. Once Planckian densities are reached, according to our prior assumption,  a quantum bounce transitions the universe back to the expanding  branch. Once more  the temperature dilutes and as it falls below  $T_c$,  the next cycle begins. One can compute the time period of these cycles just by integrating the Hubble equation (\ref{hubble}). In appendix \ref{appendix} we have obtained the expression in the approximation where we neglect non-relativistic matter contribution to the Hubble equation (\ref{hubble}) (see also~\cite{BM})~\footnote{We have ignored the time spent in the thermal bounce phase, because approximately it is given by: $\tau_b\sim M_p/T_c^2$, which is much shorter than $\tau$ as long as $T_c^4\gg\La$.}:
\be
\tau\approx {\sqrt{3}\pi M_p\over 2 \sqrt{\La}}\,.
\label{time-period}
\ee

Importantly, this is a constant and does not change from cycle to cycle as it does not depend on $\Oa_r,\Oa_m$ (As we will see shortly, the $\Oa$'s will increase from cycle to cycle with the gradual expansion of the universe.). Another way to think about this is that although the cycles are asymmetric in scale factor, it is periodic in energy densities, and it is the various energy densities that the Hubble rate is sensitive to.
\subsection{Energy Exchange \& Inflation}
Let us now turn our attention to the overall growth of the scale-factor, which is best expressed in terms of entropy growth from
cycle to cycle. The amount by which the entropy grows as matter gets converted into radiation depends on the details of the microphysical processes involved, but for us the only thing that is important to realize is that since the different $\rho$'s and $H(t)$ are periodic functions of time, this  growth in entropy density will not change from cycle to cycle. For the purpose of illustration in  appendix~\ref{appendix}, we considered a very simple scenario where in the non-thermal phase the non-relativistic species can simply decay into the radiative degrees. If the decay time is much larger than the time period of the cycle, $\Ga\tau\ll 1$, one can analytically compute the increase in the scale-factor in a given cycle:
\be
\ka\approx  {1.6\mu\Ga M_p T_c g^{1/4}\over \La^{3\over4}}+\cO(\mu^2)\ .
\ee
As expected the increase in entropy depends on the various mass scales involved, the parameter $\mu$ which controls how much of non-relativistic matter is present in the fluid, and $\Ga$ which determines how fast the massive particles decay into radiation. It is easy to see that by choosing appropriate parameter values one can make this number small.

Now, in a more realistic paradigm one will have several species involved. A stringy  model could involve a thermal Hagedorn phase  near the bounce when the massive string states are kept in thermal equilibrium with the massless degrees~\cite{emergent,BM},  but below some  critical temperature when the relevant scattering cross-sections become too small, they would fall out of equilibrium. If these modes consequently  decay into radiation,  entropy would be  produced in a manner very similar to what we discussed above. After the turnaround, as the universe starts to contract and the energy density increases, the massive states can  be replenished from radiation by scattering processes. In this case the parameter $\ka$ would capture  stringy physics involving  thermodynamics, scattering cross-sections  and decay rates.

Finally, let us comment briefly on the graceful exit mechanism. Once we replace the negative CC with a negative potential of the form depicted in figure~\ref{fig:potential}, the universe can exit the cyclic inflationary mechanism.
This is possible because, if the slope of the potential suddenly becomes steep the total scalar energy density can become positive during a single contraction phase. Once the total scalar energy density is positive, since the energy density can only increase in the contracting phase, the scalar field cannot turn back in the negative potential region (at the turning point this would imply negative total energy). This is a well known result, see for instance~\cite{linde,BM}. The universe  bounces one last time when  the energy density reaches $\sim\rho_b$. As long as the bounce occurs when  the scalar field is already in, or  is ``sufficiently near'',  the positive part of the potential, the present universe  will emerge dynamically with a positive cosmological constant after the graceful exit from the cycling inflationary  phase. Note, that the universe cannot turnaround any more as  the scalar energy density is no longer negative. Moreover, after the bounce since the kinetic energy of the scalar field starts to redshift as  $a^{-6}$, even if it dominates matter/radiation, it will  quickly become subdominant ensuring  the entry into a matter/radiation dominated universe.

Approximate calculations corroborating the above argument was presented in~\cite{BM}, and we are currently performing a detailed numerical exploration~\cite{tomi-future} of the entire parameter space to determine whether any fine-tuning is required for the mechanism to succeed.
\section{CMB Spectrum}\label{sec:perturbations}
\subsection{Scale-invariant envelope with Periodic Modulations}
Let us first provide a general argument as to why we expect to obtain a nearly scale-invariant spectrum in our model.
We are interested in tracking $\Phi_k$. As the scale-factor evolves, so does the  physical wavelength, $\la_k(t)$. Now the evolution of $\Phi_k$ at a given physical wavelength, $\la$, depends on the background environment that the mode ``sees'', \ie it depends on  the different energy densities  involved, $\{\rho_i(t)\}$, where $i$ labels  the various fluid components.  Thus another way of looking at the problem is to realize that the evolution of $\Phi_k$ depends on the curves $\{\rho_i(\la)\}$, and these curves will in general be different for different comoving modes. For simplicity, let us first consider the case when we a $-$ve CC (then cycles are precisely periodic in energy densities), and look at two modes which are related by $\ka$, the factor by which the universe expands in a given cycle:
\be
\ke=(1+\ka) k\Ra (1+\ka)\la_{\ke}=\la_k
\ee
The curves $\{\rho_i(\la)\}$   for these two modes are completely identical, the $\ke$-mode lagging behind a cycle as compared to the $k$-mode. Therefore as one evolves from $\la\ra 0$, the sub-Hubble phase, to $\la\ra \infty$, the super-Hubble phase where the fluctuations freeze, these two modes should change by the same factor.
If we use the traditional Bunch-Davis vacuum initial conditions at $t\ra -\infty$ (or $\la\ra 0$) like in standard inflation, then the sub-Hubble metric power spectrum is scale-free: $\Phi_k\ra k^{-3/2}$~\footnote{Bunch-Davis vacuum quantum fluctuations are implemented on the canonical Mukhanov variable $v_k$ which is nevertheless straight-forwardly related to $\Phi_k$. In particular the quantum vacuum fluctuations, $v_k\sim k^{-1/2}$ yields $\Phi_k\sim k^{-3/2}$, see for a review~\cite{Brandenberger}.}.
According to our previous arguments therefore, as perturbations become super-Hubble ($t\ra \infty$ and $\la\ra \infty$) this scale-invariance should be preserved, modulo periodic deviations:
\be
k^3|\Phi_k|^2=\ke^3|\Phi_{\ke}|^2\Ra\cP_{\Phi}(\ln k+\ln(1+\ka))=\cP_{\Phi}(\ln k)
\ee
Since $\ka$ is typically a small number in our model, we expect to have a near scale-invariant spectrum.
\begin{figure}[htbp]
\begin{center}
\includegraphics[width=0.35\textwidth,angle=0,scale=0.8]{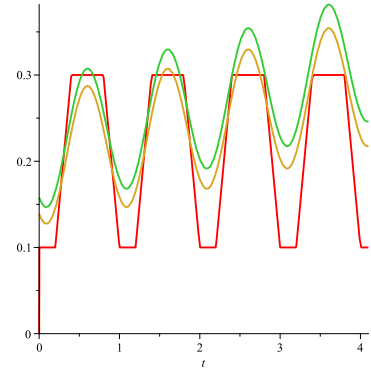}
\end{center}
\caption{The Last Exit: The curve in the red corresponds to the cosmological time-scale, while the green and the yellow curves represent the physical wavelengths of two different comoving modes. The two modes are initially in a Mixed phase but then they make their last exits after which they evolve as super Hubble modes. While the green wavelength exits in the third cycle, the yellow curve has to wait for another cycle to make it's last exit. The two modes experiences identical background cosmology, the yellow curve only lagging behind by a cycle as compared to the green curve.
  \label{fig:lastexit}}
\end{figure}

Let us actually try to calculate the spectrum by  making a few simple assumptions. Firstly, we are going to assume that as long as wavelengths of fluctuations are smaller than the cosmological time scale their power spectrum is determined by the energy density of the fluid, vis a vie (\ref{amplitude}). Secondly, we will assume that once the modes become super-Hubble they stop evolving and freeze out. These are indeed what one finds in GR and as we will see, in our scenario the freeze-out occurs near the turnaround, away from the non-GR bounce. Hence we expect these assumptions to hold true but we do want to caution the readers about possible modifications coming from the non-GR bouncing phase.

Having said that, let us go ahead and try to calculate the amplitude of perturbations, which essentially boils down to calculating the amplitude at the time of the ``last exit''. To determine when this last exit occurs let's first look at the cosmological time scale. In the usual monotonically expanding universe this is just given by the Hubble radius. However, during the bounce the time scale, $\tau_b$ is determined by the bounce energy density, $\rho_b$. Typically
\be
\tau_b=\bt{ M_p\over \sqrt{\rho_b}}
\label{taub}
\ee
where $\bt\sim \cO(1)$ constant. The cosmological time scale thus approximately behaves as follows: During the bounce phase it  stays a constant and is given by $\tau_b$, (\ref{taub}). After the bounce, radiation dominated GR takes over, the cosmological scale just corresponds to the Hubble scale, and starts to increase. Near the turnaround phase, it reaches it's maximum value   which is essentially the time period of the cycle determined by cosmological constant, see (\ref{time-period}). After the turnaround, the cosmological scale decreases back to it's minimum value near the bounce, and this cycle repeats itself, see fig.~\ref{fig:lastexit}.

In the mean time, the physical scale corresponding a particular comoving perturbation keeps oscillating with the cycles of the universe, but because of the asymmetry gradually expands. Thus, there is an initial ``pure sub-Hubble'' phase where the fluctuations are always smaller than the cosmological scale. Then there appears a ``mixed phase'' where the fluctuations go in and out of the ``Hubble radius''. Finally, there comes a last cycle where the wavelength goes out of the Hubble radius never to return, see figure~\ref{fig:lastexit}. We call this the last exit. According to the two assumptions we have made, all we have to do is to calculate the amplitude of fluctuations of the modes at their last exit point.

For this purpose let us focus on a single cycle. Consider  modes which are super-Hubble  at the bounce, \ie $\la_k\gtrsim \tau_b$. After the bounce in the radiation dominated expansion phase, the Hubble radius grows faster than the scale factor, and hence these modes start to re-enter the Hubble radius. This process continues till we reach the turnaround point when the Hubble radius stops growing and therefore the modes stop re-entering. After the turnaround in the radiation dominated contraction phase, precisely the opposite happens, \ie the modes start to exit the Hubble radius. Now, and this is the key point, since in our model the universe contracts a little less than it expands, those modes which are the first to exit the Hubble radius in the contraction phase, never come back inside the Hubble radius in the following cycle. They have just made their last exit after which their amplitudes freeze. This is illustrated in figure~\ref{fig:lastexit}.

We are now ready to calculate the spectrum. To this end we need to compute the energy density of radiation at the time the modes exit during contraction. The radiation density is, as usual,  given by
\be
\rho_r=\rho_b\LF{a\over a_b}\RF^{-4}
\ee
To obtain the spectral dependence we need to use the Hubble crossing condition
\be
\la_k=a/k=H^{-1}\Ra k\propto (-t)^{-\2}\ ,
\ee
since during the radiation dominate contraction, $H\propto 1 /t$, and $a\propto \sqrt{-t}$. Thus we have
\be
\rho_r\propto (-t)^{-2}\propto k^4
\ee
If $k_1$ corresponds to the comoving wave which is the first one to exit the Hubble radius during contraction, then
all the modes with values up to $k_1(1+\ka)$ also exits the Mixed phase in the same cycle.
Thus the power spectrum for the modes that exit the Mixed phase in a given cycle reads
\be
\cP_{\Phi}=\ga {\La\over M_p^4}\LF{k\over k_1}\RF^{4}\for k_1<k<(1+\ka)k_1
\ee
$\ga$ being an $\cO(1)$ parameter relating the Power-spectrum to the energy density of the fluid; we note that since the modes exit near the turnaround the energy density in radiation which controls the amplitude (\ref{amplitude}) must be close to $\La$. Now, once the wave number increases by a factor $(1+\ka)$, the fluctuations can no longer exit the Mixed phase, because they re-enter the Hubble radius during the following expansion. They therefore,  must wait for the next cycle for their last exit, see figure~\ref{fig:lastexit}, and consequently the power spectrum repeats.

A few remarks are in order. Although our arguments are quite general and certainly valid in GR, it is possible that the Quantum bounce regime may modify the analysis. To perform a complete treatment, one would  need to work with a  concrete model for the bounce, and then track the perturbations (most likely numerically) as they evolve through various cycles. We are currently engaged in such an analysis~\cite{tomi-future}. It is worth point out though that, given that the short wavelengths tend to decouple from cosmology it seems a reasonable assumption that they won't feel the effect of the bounce and hence their fluctuations will continue to be given by the standard expression~\ref{amplitude}. Also, it is hard to imagine that the super-Hubble constant mode in GR will disappear in the more general theories. See for instance~\cite{BKM}, for a specific higher derivative example where it was found that the super-Hubble modes indeed freeze out.

\subsection{Spectral Tilt}
The fact that the energy density of the scalar field during the inflationary phase is not a constant but slowly evolves implies that the spectral envelope (without the periodic wiggles) is not going to be precisely scale-invariant but have a tilt. Now we are working on the assumption that  the amplitude of oscillation freezes when the physical wavelength of a given comoving mode equals the cosmological time scale. This in turn depends at what scale the turnaround  occurs. In our scenario, the turnaround occurs when the positive radiation density cancels the negative scalar field density. Thus,
$\cP_{\Phi}\propto |\rho_{\phi}|$, and the power spectrum will change slowly as $\phi$ evolves.

Now, if $V'(\phi)=0$, $\phi$ would  just get Hubble damped and freeze. However, for a non-zero $V'(\phi)$ it will evolve in a manner very similar to the inflationary slow roll:
$3H_{av}\dot{\phi}=-V'(\phi)$, so that
\be
\rho_{\phi}=V(\phi)+{\dot{\phi}^2\over 2}\approx V(\phi)+{V^{'2}(\phi)\al^2M_p^2\over 18 V(\phi)\ka^2}
\ee
Depending upon the parameters of the potential it is possible for $\rho(\phi)$ to either increase or decrease as $\phi$ evolves resulting in a red or a blue tilt respectively. For instance, if our potential in the inflating region is modeled as $V(\phi)=-\La(1+e^{-\nu\phi/M_p})$, with $e^{-\nu\phi/M_p}\ll 1$, then one finds
\be
\rho_{\phi}\approx -\La[1+e^{-\nu\phi/M_p}-({\nu^2\al^2/18\ka^2})e^{-2\nu\phi/M_p}]
\ee
Since $\ka$ is a small parameter one can check that it is possible for the third term to dominate over the second so that we obtain a red tilt. We plan to investigate the detail dynamics in future, the  important point for now is that, in general we will have a non-zero  spectral tilt which can be either red or blue. Thus we expect  the power spectrum to have the following approximate form:
\ba\label{final}
\cP_{\Phi}=\ga{\rho_b\over M_p^4}\LF{k\over k_1}\RF^{\eta_s-1}f(k)
\ea
where $f(k)$ is defined as
\ba\label{f}
&f(k)&=\LF{k\over k_1}\RF^{4}\for k_1<k<(1+\ka)k_1\nonumber\\
\& &f(k(1+\ka))&=f(k)
\ea

In particular,  a distinctive feature of these modulations is that $f$ varies from 1 to $1+4\ka$ as the wave number goes from $k_1$ to $(1+\ka)k_1$. In other words, the amplitude of fluctuations, $\Da f = 2\ka$, is related to the period of fluctuations, $\Da \ln k= \ka$ in a simple way. This is important since previously very different physical models have been considered which give similar oscillatory modulations of the spectrum, but the amplitude and the period are typically two independent parameters~\cite{martin}.
\subsection{Fitting WMAP-7yr data}
To estimate the effect of the wiggles as a  good approximation to the discontinuous wiggles (\ref{f}), one can consider a smooth oscillatory modulation since it has almost the same shape and oscillatory behavior and is also more appropriate to be used to calculate the angular power spectrum using CAMB or COSMOMC. We can therefore use a power spectrum of the form:
\be\label{final}
\cP_{\Phi}={\La\over 3M_p^4}\LF{k\over k_0}\RF^{\eta_s-1}\LT1+2\ka\cos\LF{2\pi\over \ka}\ln{k\over k_0}+\ta\RF\RT
\ee
where $k_0$ is an arbitrary normalization scale. 
 The main additional parameter as compared to the standard primordial spectrum is $\ka<1/2$, although we also have an additional phase factor in  $\ta\in [0,2\pi]$.
\begin{figure}
\begin{center}
\vspace{0.1cm}
\psfig{figure=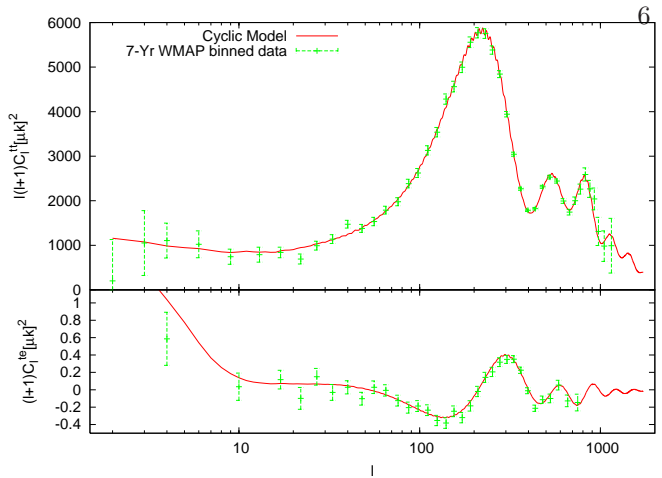,width=0.14\textwidth,angle=-90}
\vspace{1.8cm}
\end{center}
\caption{\small
Resultant best fit $C_l^{tt}$ (top panel) and $C_l^{te}$ (bottom panel) from the cyclic inflation in comparision with 7-year WMAP binned data.}
\label{fig:cl}
\end{figure}

In order to see how well the cyclic inflation fits the  CMB data, we have done some coarse sampling in the parameter space. For  baryonic abundance; $\Omega_b h^2 = 0.0228$, cold dark matter abundance, $\Omega_{dm} h^2 = 0.1156$, $h=0.70$, the optical depth, $\tau=0.082$, and parameters of the primordial spectrum of $n_s=0.963$, $\ka =0.0455$ and $\ta=2.546 (radian)$, we found a very good likelihood of $-2ln(L)=-7470.5$ to 7-year WMAP data~\cite{wmap7}.

In Fig.~\ref{fig:cl}, we see the resultant best fit angular power spectrum ($tt$ and $te$) from the cyclic model of inflation. In comparison with the best fit power-law form of the primordial spectrum and power-law with running spectral tilt model, our model can improve the likelihood by $\Delta \chi^2 = -3.9$  and $\Delta \chi^2 = -2.7$ respectively. It is interesting that in comparison with the phenomenological model of power-law with running our model has a distinctively better likelihood, although the models have similar number of degrees of freedom. Note here that we have assumed a fixed value of $k_0=0.05 {Mpc}^{-1}$ in  Eq.~(\ref{final}). In comparison with the simple power-law form of the primordial spectrum, considering the  additional free parameters in our theory, the improvement in the likelihood is not  as big as to claim that our model is preferred by the data~\cite{HSS09_PV09}, but at the same time, the improvement in the likelihood is not insignificant. Allowing tensor modes in the analysis will not improve the best fit likelihood  and we expect a very small value of $r$ (tensor to scaler ratio). We will leave a proper cosmological parameter estimation using CMB and other cosmological data for future
publication, but the derived likelihood and the best fit parameters already suggests that this model should provide a good fit.
\section{The Standard Cosmological Puzzles and Future Outlook}\label{sec:conclusions}
Before we conclude, let us look at the problem of growth of inhomogeneity/overproduction of black holes.  The matter fluctuations, $\da \equiv \da \rho/\rho$ can only grow as long as their wavelengths are larger than the Jean's length, $\la_J$,  given by: $ \la_J\sim c_s{M_p/\sqrt{\rho}}\mx{ where } c_s^2\equiv {\p p/\p \rho}$, is the sound velocity square. Now, in our scenario the cycles are short, and most of the energy content is in radiation so that the sound speed is very close to the speed of light, and accordingly  $\la_J \sim H^{-1}\sim \tau$. In other words, the sub-Hubble fluctuations don't grow.   On the other hand, once the wavelengths become  larger than the cosmological time scale, $\tau$, they become super-Hubble fluctuations and evolves according to the Poisson equation: $\da _k= {k^2\Phi_k/ (a^2\rho)}$,
where $\Phi_k$ is the Newtonian potential characterizing the metric perturbations. Now in the super-Hubble phase, $\Phi_k$ becomes a constant while $\rho$ oscillates between a minimum and a maximum energy density. Thus we have: $\da _k< {k^2\Phi_k/( a^2\rho_{\min})}\sim  {k^2\Phi_kM_p/( a^2\sqrt{\La})}$. In other words $\da _k$ falls as $a^{-2}$ in the super Hubble phase just as in ordinary inflation. Finally,  cyclic cosmologies are notoriously plagued with Mixmaster chaotic behavior as one approaches the Big Crunch/Bounce, since  anisotropies grow as $\sim a^{-6}$. However, in our model the same reasoning that resolves the flatness problem, also solves this issue. Once the cyclic-inflationary phase is ``activated'' in a small patch of the universe the chaotic Mixmaster behavior is avoided in subsequent cycles as the universe becomes more and more isotropic due to the overall growth of the scale factor.

To summarize we have presented a unique way of realizing inflation within cyclic universe scenario, where the particle trajectory is geodesically complete in either past and future. Although in each cycle the universe expands only a little bit, one can obtain a large number of total inflationary e-foldings, $\cN_{\mt{tot}}$ over many  many cycles.  As long as $\cN_{\mt{tot}}\gtrsim 60$  it is clear that we can explain the horizon and the flatness problems.  Furthermore, every cycle leave their quantum imprint in the CMB with potentially observable
wiggles determined by the enhancement of the scale factor during every oscillation.

We should however emphasize that there are still several aspects of the model which needs a more comprehensive numerical investigation. For instance, what are the conditions necessary to maintain thermal equilibrium during the bounce phase? (In fact, all we really require is that thermal equilibrium be established at some point of time in every cycle, that is sufficient to guarantee the periodicity in energy density.) What about the graceful exit problem. It is clear that whether we can transition from the inflationary phase to the positive potential region depends on various parameters of the model, do these parameters need to fine-tuned? Finally, what effect does the bounce play in the evolution of the perturbations. These are important questions and we are currently pursuing them using numerical simulations~\cite{tomi-future}. Finally, it is also foreseeable that our scenario can potentially  enhance  non-gaussianity through stochastic resonance during these oscillations, again an interesting future direction to pursue.

 AM and AS acknowledge the support of the EU Network UniverseNet (MRTN-CT-2006-035863) and thanks Jan Hamann and Xingang Chen for useful discussions.
\section{Appendix: $\tau$ and $\Da S$}\label{appendix}

Here we calculate the approximate time period and the increase in entropy in a given cycle. We will calculate this under the approximation that the dust component can be neglected as $\mu\ll 1$ and treat $\Oa$ as a constant as the amount of matter decay in a given cycle is negligible.

To calculate the time period we start by rewriting the Hubble equation:
\be
dt={da\over\dot{a}} = {\sqrt{3}M_p da\over T_c^2a\sqrt{{\Oa_r\over a^4}-{\La\over T_c^4}}}
\ee
so that the time period is given by
\be
\tau\approx {2\sqrt{3}M_p \over T_c^2}\int_{a_c}^{a_T} {da \over a\sqrt{{\Oa_r\over a^4}-{\La\over T_c^4}}}
\ee
where we have neglected the duration in the thermal bounce phase as it is going to be much shorter than the above integral.
Now  the turnaround scale factor is given by
\be
a_T=\LF{\Oa_r T_c^4\over \La}\RF^{1/4}
\ee
Thus the above integral can be re-expressed as
\be
\tau= {2\sqrt{3}M_p \over \sqrt{\La}}\int_{a_c}^{a_T} {da \over a\sqrt{\LF{a_T\over a}\RF^4-1}}= {2\sqrt{3}M_p \over \sqrt{\La}}\int_{\en}^{1} {y dy \over \sqrt{1-y^4}}
\ee
where $\en\equiv a_c/a_T\ll 1$.
Thus we have
\be
\tau\approx {2\sqrt{3}M_p \over \sqrt{\La}}\int_{0}^{1} {y dy \over \sqrt{1-y^4}}= {\sqrt{3}\pi M_p \over 2\sqrt{\La}}
\ee

In a similar manner we can proceed to calculate the approximate entropy increase. Phenomenologically,  energy exchanges can be captured  by generalizing  conservation equations~\cite{barrow,dabrowski,emergent} for the two fluids to:
\be
\dot{\rho_r}+4H\rho_r=T_c^4 s\,,~~~~~~~~\dot{\rho_m}+3H\rho_m=-T_c^4s
\label{continuity}
\ee
which now includes an energy exchange term. We can easily compute the net entropy increase~\cite{BM}:
\be
\dot{S}=\dot{S_r}+\dot{S_m}=a^3s\LF{3b_rT_c\over 4\rho_r^{1/4}}-{T_c\over M}\RF=a^3s\LF{T_c\over T_r}-1\RF
\label{entropy-growth}
\ee
where we have used the usual expressions for the thermodynamic entropies associated with matter and radiation:
\be
S_r={4\rho_r V\over 3T}={4\over 3}g^{1/4} \Oa_r^{3/4}\,,~~~~S_m={\rho_m V\over M} =\Oa_m\,.
\label{entropy-formula}
\ee
Here  $M\approx T_c$ corresponds to the mass of the non-relativistic particles, and in our convention,  $g=(\pi^2/30) g_{\ast}$, where $g_{\ast}$ is the number of ``effective'' massless degrees of freedom.

The  growth of the entropy in a given cycle will obviously depend on the phenomenological function $s$. For the simplest case, which captures the physics of the decay of the non-relativistic species into radiation, $s$ is given by
\be
s={\Ga \rho_m\over T_c^4}\,,
\label{s}
\ee
$\Ga$ being the decay rate. One can now obtain the entropy increase in a given cycle by simultaneously solving for the Hubble equation (\ref{hubble}), the continuity equations (\ref{continuity}) and entropy growth (\ref{entropy-growth}). For some special cases  this can also be done analytically. For the purpose of illustration we are going to  look at the case when the decay time is much larger than the time period of the cycle, $\Ga\tau\ll 1$. This means that basically the $\Oa$'s change very little over a given cycle and for the purpose of estimating the entropy increase from (\ref{entropy-formula}) we can treat them to be constants. We can approximate $e^{-\Ga t}\approx 1$ in the expression for $\dot{S}$. Using (\ref{entropy-growth}), (\ref{hubble}), (\ref{s}) and (\ref{energydensities}) we find
\be
{dS\over da}={\sqrt{3}\Ga \Oa_m M_p\over a\sqrt{\La}}{\LF{a\over a_c}-1\RF\over  \sqrt{\LF{a_T\over a}\RF^4-1}}
\ee
Thus the increase in entropy in a given cycle can be calculated as
$$
\Da S={2\sqrt{3}\Ga \Oa_m M_p\over \sqrt{\La}}\int_{a_c}^{a_T} {da\over a}{\LF{a\over a_c}-1\RF\over a \sqrt{\LF{a_T\over a}\RF^4-1}}$$
$$={2\sqrt{3}\Ga \Oa_m M_p\over \sqrt{\La} }\int_{y_c}^{1} dy{ y\LF {y\over y_c}-1\RF \over \sqrt{1-y^4}}$$
$$\approx {2\sqrt{3}\Ga \Oa_m M_p\over \sqrt{\La}y_c }\int_{0}^{1} dy{y^2 \over \sqrt{1-y^4}}
$$

Now, we know that at the transition temperature the energy density in radiation must be given by
\be
\rho_r=gT_c^4=T_c^4{\Oa_r\over a_c^4}\Ra a_c=\LF{\Oa_r\over g}\RF^{1/4}
\ee
Using the expression for $a_T$ then we find
\be
y_c={a_c\over a_T}={\La^{\4}\over g^{\4}T_c}
\ee
Using the expressions for energy density and entropy we can also easily relate $\Oa_m$ with the total entropy $S$:
\be
\Oa_m\approx {3\mu S\over 4}
\ee
Thus we finally have
\be
\Da S\approx {3\sqrt{3}\mu\Ga S M_p\over 2\sqrt{\La}y_c }\int_{0}^{1} dy{y^2 \over \sqrt{1-y^4}}\approx {1.6\mu\Ga S M_p T_c g^{\4}\over \La^{3\over4}}
\ee



\begin{thebibliography}{99}

\bibitem{wmap7}
E.~Komatsu et. al,  [arXiv:1001.4538 [astro-ph]]

\bibitem{RM}
  A.~Mazumdar and J.~Rocher,
  arXiv:1001.0993 [hep-ph].




\bibitem{Linde1}
A.~D.~Linde, D.~A.~Linde and A.~Mezhlumian,
  Phys.\ Rev.\  D {\bf 49}, 1783 (1994)
A.~Borde and A.~Vilenkin,
  Phys.\ Rev.\ Lett.\  {\bf 72}, 3305 (1994)
[arXiv:gr-qc/9312022].
A.~Borde, A.~H.~Guth and A.~Vilenkin,
  Phys.\ Rev.\ Lett.\  {\bf 90}, 151301 (2003)
  [arXiv:gr-qc/0110012].
 T.~Vachaspati and M.~Trodden,
  Phys.\ Rev.\  D {\bf 61}, 023502 (1999)
  [arXiv:gr-qc/9811037].

\bibitem{tolman} R.C.Tolman,
Phys.\ Rev.\ {\bf 37}, 1639  (1931)

\bibitem{narlikar}
H.~Bondi and T.~Gold,
  Mon.\ Not.\ Roy.\ Astron.\ Soc.\  {\bf 108}, 252 (1948).
F.~Hoyle, MNRAS, {\bf 108}, 372 (1948);
  arXiv:0801.2965 [astro-ph].

\bibitem{Starobinsky}
  A.~A.~Starobinsky,
  Phys.\ Lett.\  B {\bf 91} (1980) 99.



\bibitem{ekcyclic}
J.~Khoury, B.~A.~Ovrut, P.~J.~Steinhardt and N.~Turok,
  Phys.\ Rev.\  D {\bf 64}, 123522 (2001)
  [arXiv:hep-th/0103239];
P.~J.~Steinhardt and N.~Turok,
  Phys.\ Rev.\  D {\bf 65}, 126003 (2002)
  [arXiv:hep-th/0111098];
  Science {\bf 296}, 1436 (2002).

\bibitem{barrow}
  J.~D.~Barrow, D.~Kimberly and J.~Magueijo,
  Class.\ Quant.\ Grav.\  {\bf 21}, 4289 (2004)
  [arXiv:astro-ph/0406369];
 J.~D.~Barrow and M.~P.~Dabrowski,
MNRAS {\bf 275},  850 (1995).

\bibitem{phantom}
  M.~G.~Brown, K.~Freese and W.~H.~Kinney,
  JCAP {\bf 0803}, 002 (2008)
  [arXiv:astro-ph/0405353];
   L.~Baum and P.~H.~Frampton,
  Phys.\ Rev.\ Lett.\  {\bf 98}, 071301 (2007)
  [arXiv:hep-th/0610213];

 \bibitem{emergent}
  T.~Biswas,
  arXiv:0801.1315 [hep-th];
  T.~Biswas and S.~Alexander,
  arXiv:0812.3182 [hep-th].
\bibitem{peter}
  F.~T.~Falciano, M.~Lilley and P.~Peter,
  Phys.\ Rev.\  D {\bf 77}, 083513 (2008)
  [arXiv:0802.1196 [gr-qc]].

\bibitem{BM}
  T.~Biswas and A.~Mazumdar,
  Phys.\ Rev.\  D {\bf 80}, 023519 (2009)
  [arXiv:0901.4930 [hep-th]].
\bibitem{saridakis}   Y.~F.~Cai and E.~N.~Saridakis,
  JCAP {\bf 0910}, 020 (2009)
  [arXiv:0906.1789 [hep-th]].


\bibitem{robert}
  R.~H.~Brandenberger,
  Phys.\ Rev.\  D {\bf 80}, 023535 (2009)
  [arXiv:0905.1514 [hep-th]];
   J.~Zhang, Z.~G.~Liu and Y.~S.~Piao,
  arXiv:1007.2498 [hep-th].
Y.~S.~Piao,
  Phys.\ Lett.\  B {\bf 691}, 225 (2010)
  [arXiv:1001.0631 [hep-th]].




\bibitem{justin}
  E.~I.~Buchbinder, J.~Khoury and B.~A.~Ovrut,
  Phys.\ Rev.\  D {\bf 76}, 123503 (2007)
  [arXiv:hep-th/0702154].
   J.~L.~Lehners, P.~McFadden, N.~Turok and P.~J.~Steinhardt,
  Phys.\ Rev.\  D {\bf 76}, 103501 (2007)
  [arXiv:hep-th/0702153].

J.~Khoury and P.~J.~Steinhardt,
  Phys.\ Rev.\ Lett.\  {\bf 104}, 091301 (2010)
  [arXiv:0910.2230 [hep-th]].

\bibitem{BMS}
 T.~Biswas, A.~Mazumdar and W.~Siegel,
  JCAP {\bf 0603}, 009 (2006)
  [arXiv:hep-th/0508194].
 T.~Biswas, R.~Brandenberger, A.~Mazumdar and W.~Siegel,
  JCAP {\bf 0712}, 011 (2007)
  [arXiv:hep-th/0610274].
  \bibitem{matter}
  Y.~F.~Cai, T.~t.~Qiu, R.~Brandenberger and X.~m.~Zhang,
  Phys.\ Rev.\  D {\bf 80}, 023511 (2009)
  [arXiv:0810.4677 [hep-th]].
 \bibitem{landscape}
M.~R.~Douglas and S.~Kachru,
  Rev.\ Mod.\ Phys.\  {\bf 79}, 733 (2007)
 [arXiv:hep-th/0610102];
  H.~Firouzjahi, S.~Sarangi and S.~H.~H.~Tye,
  JHEP {\bf 0409}, 060 (2004)
  [arXiv:hep-th/0406107].
\bibitem{linde}
  G.~N.~Felder, A.~V.~Frolov, L.~Kofman and A.~V.~Linde,
  Phys.\ Rev.\  D {\bf 66}, 023507 (2002)
  [arXiv:hep-th/0202017].
\bibitem{BKM}
  T.~Biswas, T.~Koivisto and A.~Mazumdar,
  arXiv:1005.0590 [hep-th].
\bibitem{loop}
M.~Bojowald,
  Phys.\ Rev.\ Lett.\  {\bf 86}, 5227 (2001)
  [arXiv:gr-qc/0102069].
A.~Ashtekar, T.~Pawlowski, P.~Singh and K.~Vandersloot,
  Phys.\ Rev.\  D {\bf 75}, 024035 (2007)
  [arXiv:gr-qc/0612104];
  A.~Ashtekar, T.~Pawlowski and P.~Singh,
  Phys.\ Rev.\ Lett.\  {\bf 96}, 141301 (2006)
  [arXiv:gr-qc/0602086];
  Phys.\ Rev.\  D {\bf 74}, 084003 (2006)
  [arXiv:gr-qc/0607039].
  G.~Date and G.~M.~Hossain,
  Phys.\ Rev.\ Lett.\  {\bf 94}, 011302 (2005)
  [arXiv:gr-qc/0407074].
  \bibitem{Brandenberger}
 V.~F.~Mukhanov, H.~A.~Feldman and R.~H.~Brandenberger,
  Phys.\ Rept.\  {\bf 215}, 203 (1992).
  \bibitem{tomi-future} T.~Biswas, T.~Koivisto and A.~Mazumdar, in progress.
\bibitem{martin}
  J.~Martin and C.~Ringeval,
  Phys.\ Rev.\  D {\bf 69}, 083515 (2004)
  [arXiv:astro-ph/0310382];
 R.~Flauger, L.~McAllister, E.~Pajer, A.~Westphal and G.~Xu,
  arXiv:0907.2916 [hep-th].
\bibitem{HSS09_PV09}
H.~Peiris and L.~Verde, Phys.\ Rev.\  D. {\bf 81}, 021302 (2010); J.~Hamann, A.~Shafieloo and T.~Souradepp, [arXiv:0912.2728 [astro-ph]]

\end{thebibliography}
\end{document}